\pdfoutput=1

\documentclass[final,onefignum,onetabnum]{siuro210301}
\usepackage{lipsum}
\usepackage{amsmath,amsfonts,amssymb}
\usepackage{graphicx}
\usepackage{epstopdf}
\usepackage{algorithmic}

\ifpdf
  \DeclareGraphicsExtensions{.eps,.pdf,.png,.jpg}
\else
  \DeclareGraphicsExtensions{.eps}
\fi

\usepackage{tikz}
\usetikzlibrary{arrows.meta, quotes}
\usetikzlibrary{arrows,decorations.markings}
\usetikzlibrary{positioning}
\usepackage{pgfplots}
 \usepackage{multirow}

\usetikzlibrary{shadows}

\usepackage{mathtools}
\usepackage{wrapfig}
\usepackage{subcaption}
\usepackage{bm}
\newcommand{\vu}{{\bf u}}
\newcommand{\xv}{{\bf x}}
\newcommand{\vk}{{\bf k}}
\newcommand{\vv}{{\bf v}}
\newcommand{\Ra}{ \textrm{Ra} }
\newcommand{\Pra}{ \textrm{Pr} }
\newcommand{\Nus}{ \textrm{Nu} }
\newcommand{\NuL}{ \Nus_L }
\newcommand{\parD}[2]{\frac{\partial #1}{\partial #2}}
 
\usepackage{tikz,pgfplots}
\usepackage{pgfplotstable}
\pgfplotsset{compat=1.6}

\usepackage{mathtools}
\newenvironment{brsm}{
  \bigl[ \begin{smallmatrix} }{%
  \end{smallmatrix} \bigr]} 

\usepackage{ulem}

\usepackage{enumitem}
\setlist[enumerate]{leftmargin=.5in}
\setlist[itemize]{leftmargin=.5in}

\newsiamremark{remark}{Remark}
\newsiamremark{hypothesis}{Hypothesis}
\crefname{hypothesis}{Hypothesis}{Hypotheses}
\newsiamthm{claim}{Claim}

\headers{A numerical investigation of Rayleigh-Bénard convection with an obstruction}{H. Mhina, S. Souley Hassane}

\title{A numerical investigation of Rayleigh-Bénard convection with an obstruction}

\author{Harieth Mhina\thanks{Trinity College, Hartford, CT. 
}
\and Samira Souley Hassane\footnotemark[2]}

\dedication{\small\textit{Project advisor: Dr. Matthew McCurdy \thanks{Department of Mathematics, Trinity College, Hartford, CT.}}}

\usepackage{amsopn}

\ifpdf
\hypersetup{
  pdftitle={A numerical investigation of Rayleigh-Bénard convection with an obstruction},
  pdfauthor={H. Mhina, S. Souley Hassane}
}
\fi

\begin{document}

\maketitle

\begin{abstract}
The phenomenon of convection is found in a wide variety of settings on different scales-- from applications in the cooling technology of laptops to heating water on a stove, and from the movement of ocean currents to describing astrophysical events with the convective zones of stars. Given its importance in these diverse areas, the process of convection has been the focus of many research studies over the past two centuries. However, much less research has been conducted on how the presence of an obstruction in the flow can impact convection. In this work, we find that the presence of an obstruction can greatly affect convection. We note occurrences where the presence of an obstruction yields similar behavior to flow without an obstruction. Additionally, we find cases with markedly different features in comparison to their counterpart without an obstruction-- notably, exhibiting long-term periodic behavior instead of achieving a constant steady-state, or the formation of convection cells versus an absence of them.
\end{abstract}

\section{Introduction}
\begin{figure} 
\centering
\definecolor{darkblue}{rgb}{0.2,0.2,0.6}
\definecolor{darkred}{rgb}{0.6,0.1,0.1}
\definecolor{darkgreen}{rgb}{0.2,0.6,0.2}

\def\arrow{
  (10.75:1.1) -- (6.5:1) arc (6.25:120:1) [rounded corners=0.5] --
  (120:0.9) [rounded corners=1] -- (130:1.1) [rounded corners=0.5] --
  (120:1.3) [sharp corners] -- (120:1.2) arc (120:5.25:1.2)
  [rounded corners=1] -- (10.75:1.1) -- (6.5:1) -- cycle
}

\tikzset{
  ashadow/.style={opacity=0},
}

\def\arrows[#1]{         
  \begin{scope}[scale=#1]
    \draw[color=darkred, right color=red, left color=red!60, %
    drop shadow={ashadow, color=red!60!black}] [rotate=225] \arrow;
    \draw[color=darkblue, right color=blue, left color=blue!60, %
    drop shadow={ashadow, color=blue!60!black}] [rotate=45] \arrow;
  \end{scope}
}

\begin{tikzpicture}
[
    auto,
     Arr/.style={-{Triangle},-{Latex[length=2.5mm]},{line width=2pt}},
                        ]
    \arrows[1];
    \node[fill=none] at (0,-0.5) {warmer};
    \node[fill=none] at (0,0.5) {cooler};
    \draw[black, line width=1.75pt] (-1.53,-1.5) -- (1.53,-1.5);
	\draw[black, line width=1.75pt] (1.5,-1.5) -- (1.5,1.5);
	\draw[black, line width=1.75pt] (1.53,1.5) -- (-1.53,1.5);
	\draw[black, line width=1.75pt] (-1.5,1.5) -- (-1.5,-1.5);
	 \node (top_node)  at (1.9,.7) { };
   	 \node (bottom_node) at (1.9,-.7) { };
    \draw[Arr] (top_node)   to [bend left=00,"{\bf g}"] (bottom_node);
\end{tikzpicture} 
 \caption{Schematic of a Rayleigh-B\'enard convection cell. \label{fig:RBCschematic}}
\end{figure}
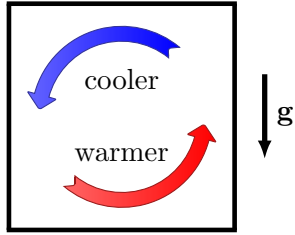

Convection is the process of moving thermal energy in a fluid as a result of a temperature difference, with natural convection defined as fluid movement due to changing densities in the fluid, or a buoyancy force. These kinds of convection arise from properties innate to the fluid-- like density-- in contrast to forced convection where some external source, like a fan, is responsible for fluid movement. With many cases of natural convection, a layer of fluid is heated from below and a temperature difference is established. The fluid at the bottom becomes less dense than the fluid at the top, which gives rises to a top-heavy arrangement that is potentially unstable, as shown with the schematic in Figure \ref{fig:RBCschematic}. Due to this instability, the fluid will tend to redistribute itself to remedy the weakness in its arrangement, resulting in circular movement of the fluid, \cite{chandrasekhar1981hydrodynamic}. This convection, now known as Rayleigh-Bénard convection, has been the focus of much research in various fields due to its importance. For a more thorough overview of Rayleigh-Bénard convection, we recommend the following references:  \cite{chandrasekhar1981hydrodynamic,getling1998rayleigh,koschmieder1993benard,lappa2009thermal}.

The first experimental studies to investigate thermal instability of fluids were done by Henri Bénard around 1900 where he noticed the formation of hexagonal cells in a thin layer of fluid heated from below, \cite{benard1900etude, benard1900mouvements, benard1900tourbillons, benard1901tourbillons, benard1901paper}. Years later in the seminal work \cite{rayleigh1916}, Lord Rayleigh showed that convection occurs only when a nondimensional quantity,
\begin{align*}
    \frac{g\alpha \beta}{k\nu}h^4
\end{align*}
(now known as the Rayleigh number) is above a threshold value, where $g$ is the acceleration due to gravity, $\alpha$ is the coefficient of volume expansion, $\beta = \lvert dT/dz \rvert$ is the uniform temperature gradient maintained between the top and bottom plates, with $h$, $k$ and $\nu$ as the height of the fluid, coefficient of thermometric conductivity, and kinematic viscosity, respectively. Later, in \cite{pellew1940maintained}, Pellew and Southwell returned to the experimental observations and theoretical results to create a more generalized argument for determining when convection would occur and calculated the critical Rayleigh number of $\Ra^*\approx1707.8$ for an infinite layer of fluid-- a value confirmed by Reid and Harris in \cite{reid1958some} by a complementary approach to the problem. For an enclosure of finite width (in comparison to the theoretical investigations of fluid in an infinitely-wide layer), Mizushima and Adachi determined the critical Rayleigh numbers for various aspect ratios, different combinations of boundary conditions, and settings with other physical considerations, like a box on a slant, \cite{adachi1996stability,mizushima1995onset,mizushima1995structural,mizushima1997sequential}.

The Rayleigh-B\'enard convection problem has multiple applications in industrial engineering such as the cooling of buildings \cite{time_periodic_cooling} and electronic devices \cite{TURAN201283} as well as in turbulent heat transfer such as in geophysics and astrophysics \cite{van_der_poel_stevens_lohse_2013}. Furthermore, Rayleigh-B\'enard convection can be used to investigate nonlinear systems as well as chaotic dynamics. Numerical investigations into Rayleigh-B\'enard convection of simple two-dimensional domains are frequent in published literature. However, there is much less research exploring convection with obstructions in the flow.

The consideration of obstructions in a flow is important in a variety of settings: in geophysical settings with rivers flowing around large rocks, in aircraft design when trying to minimize resistance from the surrounding air when in flight, or even in determining the placement of wires in laptop hardware to enhance circulation and prevent overheating. These examples assume the object in the flow to be impermeable, and therefore, the fluid must react to navigate around the obstruction. The fluid's reaction often results in different behavior than flows in settings without an obstruction. 

The purpose of this note is to numerically investigate the effect of an obstruction on the convection of an incompressible fluid inside a square enclosure that is heated from below and cooled from above. We first consider convection inside a square domain without an obstruction. This allows us to use previous literature to validate our numerical methods. Previous work of Ouertatani {\it et al.} \cite{ouertatani2008numerical} on Rayleigh-B\'enard convection in a square enclosure  provides the main quantitative benchmarks for comparison. Their study uses a finite volume formulation to conduct simulations, while our numerical simulations are computed with a finite element method (FEM) to approximate solutions to the weak form of the system of partial differential equations (PDEs). 

After we establish that our numerical methods work as expected, we address how convection is affected by an obstruction in the flow. For this end, we look at two quantitative benchmarks: the Nusselt number that evaluates vertical flux due to convection, and a mathematical energy that describes how much the system has deviated from the original state. We find that the presence of an obstruction can drastically affect flow profiles. We note occurrences where the presence of an obstruction yields similar behavior to flow without an obstruction. However, we also find cases with extremely different features in comparison to flows without an obstruction-- notably, exhibiting long-term periodic behavior instead of achieving a constant steady-state, or the formation of convection cells versus an absence of them.

\section{Problem formulation}
The Navier-Stokes-Boussinesq equations express conservation of momentum and mass, as well as how temperature affects the body force term. We pair this with the customary incompressibility condition and the advection-diffusion equation to describe how heat and the fluid interact and impact each other. These three equations make up our system, and are presented below as a system of nondimensionalized partial differential equations. We use the same nondimensional scalings from \cite{ouertatani2008numerical}, allowing us to use their results as benchmark for our solutions. 
The scalings-- also noted in \cite{lee2016three,miroshnichenko2018turbulent,zhao2006conjugate}, among others-- have nondimensional quantities denoted by a tilde:
\begin{align*}
\textbf{x} = \tilde{\textbf{x}}\,H,\,\quad \vu = \tilde{\vu} \,u_0,\,\quad t = \tilde{t}\,\frac{H}{u_0}\,,\quad \textrm{and }\quad p = \tilde{p} \,\rho_0\, u_0^2\,,
\end{align*} where $u_0=\left[g\beta(T_H - T_C)H\right]^{1/2}$ is the reference velocity, $\rho_0$ is the reference density, $\beta$ is the coefficient of thermal expansion, $g$ is the gravitational acceleration, $T_H$ and $T_C$ are the temperatures of the hot and cold plates, respectively, and $H$ is the reference length, taken to be the width of the domain for our simulations. After dropping the tildes, our system is then: 
\begin{align}
\begin{cases}
\frac{\partial \vu}{\partial t} +(\vu \cdot \nabla) \vu = -\nabla p + \left(\frac{\Pra}{\Ra}\right)^{\frac{1}{2}} \nabla^2 \vu + \theta \vk , \\[5pt]
\nabla \cdot {\vu} = 0 ,\\[5pt]
\frac{\partial \theta}{\partial t} + \vu\cdot\nabla \theta = \frac{1}{(\Ra \Pra)^{\frac{1}{2}}} \nabla^2 \theta,\\[5pt]
\end{cases} 
\end{align}
where the velocity $\vu=\begin{brsm}u\\v \end{brsm}$, pressure $p$, and temperature $\theta$ are variables. The vector $\bf{k} = \begin{brsm}0\\1 \end{brsm}$ is the unit normal vector which points in the opposite as the gravitational force and helps detail the effect of buoyancy in the system.  The Prandtl number $\Pra$ is a dimensionless ratio of momentum diffusivity to thermal diffusivity and the Rayleigh number $\Ra$ is a nondimensional constant describing the heat difference between the top and bottom surfaces of the domain. The Rayleigh number characterizes heat transfer in natural convection. When $\Ra$ is below its critical value (determined to be $\Ra^*=2585.02$ by Mizushima \cite{mizushima1995onset} for the same boundary conditions we consider), convection does not occur and heat is transferred through conduction only. When the Rayleigh number is above its critical value though, heat is transferred through convection and fluid motion begins to occur. The Prandlt number and Rayleigh number are defined respectively as:
\begin{equation*}
     \Pra =\frac{\nu}{\alpha}\,, \quad
	\Ra = \frac{g\beta(T_H - T_C)H^3}{\alpha \nu}\,,
\end{equation*}
where $\nu$ is the kinematic viscosity and $\alpha$ is the thermal diffusivity. 

The domain we consider is a square of unit area, with $x\in [-.5, .5]$ and $z\in[0,1]$. While we use $H=1$ (where the depth and width of the enclosure are the same) throughout this work, Mizushima considers convection in finite two-dimensional boxes with different aspect ratios, \cite{mizushima1995onset}.

To make sure the system is well-posed, each variable in the system has a condition it satisfies at each boundary of the domain. The temperature of the top and bottom plates are held constant at $\theta = -\frac{1}{2}$ and $\theta = \frac{1}{2}$, respectively. At the left and right sides of the domain, we have no-flux conditions, or insulating boundaries, with $\nabla \theta\cdot{\bf n}= 0$ where ${\bf n}$ is the outward pointing unit normal vector. For the velocity, we impose the no-slip condition, $\vu = {\bf 0}$ along each side of the domain; i.e., the velocity is zero at each boundary.

The initial conditions for our simulations come from the conductive state-- a velocity of zero, $\vu=\bf{0}$ and a linear temperature profile of $\theta(x,z)=0.5-z$, which also satisfies our boundary conditions. The conductive profile is a stable steady-state for $\Ra$ below a critical value and unstable for $\Ra$ above its critical value. Therefore, when we perturb this steady-state at the beginning of the simulations for Rayleigh numbers above the critical value, hotter fluid from the bottom will rise due to the buoyancy force, while the cooler fluid from the top will sink. This fluid movement forms the convection cells and helps define the preferred stable state (for non-turbulent parameter regimes).

When we consider an obstruction in the flow, we assume the obstruction is impermeable and insulated, so we impose no-slip and no-flux boundary conditions for velocity and temperature along the boundary of the obstruction. Physically, this could be representing the placement of insulated wires in an empty air channel in laptop hardware, investigated to determine optimal wire placement to enhance circulation and prevent overheating. We run simulations with different placements of the obstruction, with two placements shown in Figure \ref{fig:mesh} with a schematic of mesh discretizations. Other works focused on convection with obstructions often consider rectangular obstructions (usually where the obstruction generates a heat source) \cite{miroshnichenko2018turbulent,paroncini2009natural,zhao2006conjugate}. Investigations which consider circular obstructions (or cylindrical in three dimensional studies) where the obstruction is providing heat \cite{ghaddar1992natural,lee2016three} or in enclosures with isothermal boundary conditions along all borders and the obstruction \cite{kim2008numerical,lee2010natural}. Although these each consider similar ideas of convection with an obstruction, an exact comparison of results is not possible since we consider different boundary conditions and do not allow the obstruction to conduct heat.

Next, we define three quantitative measures of convection. The first is the Nusselt number, the physically-relevant measure of the ratio of convective to conductive heat transfer,
\begin{equation}\label{eqn:nuss}
\Nus(t) = 1 + \int_{\Omega} \theta\,\vu\cdot{\bf k}\,d\Omega\,,
\end{equation} as defined in \cite{doering1996variational,doering2001upper,howard1963heat,howard1972bounds}. Second, we define a mathematical energy, 
\begin{equation*}
    2\,E(t)=\frac{1}{\Pra}\int_{\Omega}\lvert\vu \rvert^2 d\Omega+ \int_{\Omega}(\theta - \theta_i)d\Omega\,,
\end{equation*}
where $\theta$ is the temperature, $\theta_i$ is the initial temperature profile, $\vu$ is the velocity, and $\Pra$ is the Prandtl number. This is a measure of the deviation between the flow and the conductive state, and is used in analytical arguments to determine the critical Rayleigh number needed for the transition from $\frac{dE}{dt}<0$ to $\frac{dE}{dt}>0$, which notes a transition to convection. One recent example of this kind of analysis studied convection in a  superposed fluid-porous medium, \cite{mccurdy2019convection}. Lastly, we define the local Nusselt number along the lower boundary with \begin{equation*}
    \NuL(x) = -\frac{\partial \theta}{\partial z} \Big | _{z=0}\,.
\end{equation*} The two definitions of the Nusselt numbers are equivalent with some mathematical manipulation-- by applying the divergence theorem and boundary conditions-- as noted in some of the pioneering works on analysis of Rayleigh-B\'enard problem, \cite{howard1963heat,howard1972bounds}.

To compare our results to the work done by Ouertatani {\it et al.}, we consider three Rayleigh numbers $\Ra=10^4,\,10^5,\,10^6$ in our simulations. We begin our investigation at $\textrm{Ra} =10^4$ since no convection is observed at $\textrm{Ra} = 10^3$, as it is below the critical Rayleigh number of $\textrm{Ra}^*\approx2585$. As $\Ra$ increases, the physical heat difference between the top and bottom of the domain increases and fluid motion within the domain becomes more vigorous. As a result, we expect the velocity of the fluid to increase and the temperature field to reflect this increase by forming more defined convection cells. Moreover, with our quantitative measures of convection, we expect the Nusselt number and energy to increase as the Rayleigh number increases since more heat is being pumped into the system.

\section{Numerical methods}
Since exact analytical solutions are, in general, impossible to find, we use a Finite Element Method (FEM) to approximate solutions to our system. The unit square domain is discretized and divided into a finite number of small triangular elements  for all three different domains. Figure \ref{fig:mesh} below shows the schematic of the meshes for the domains. Our simulations are run on a more refined grid to allow for more accurate solutions to be computed; we use a length scale of 128, in comparison to the length scale of 32 shown in the schematic.
\begin{figure}[]
    \centering
    \includegraphics[width=\textwidth]{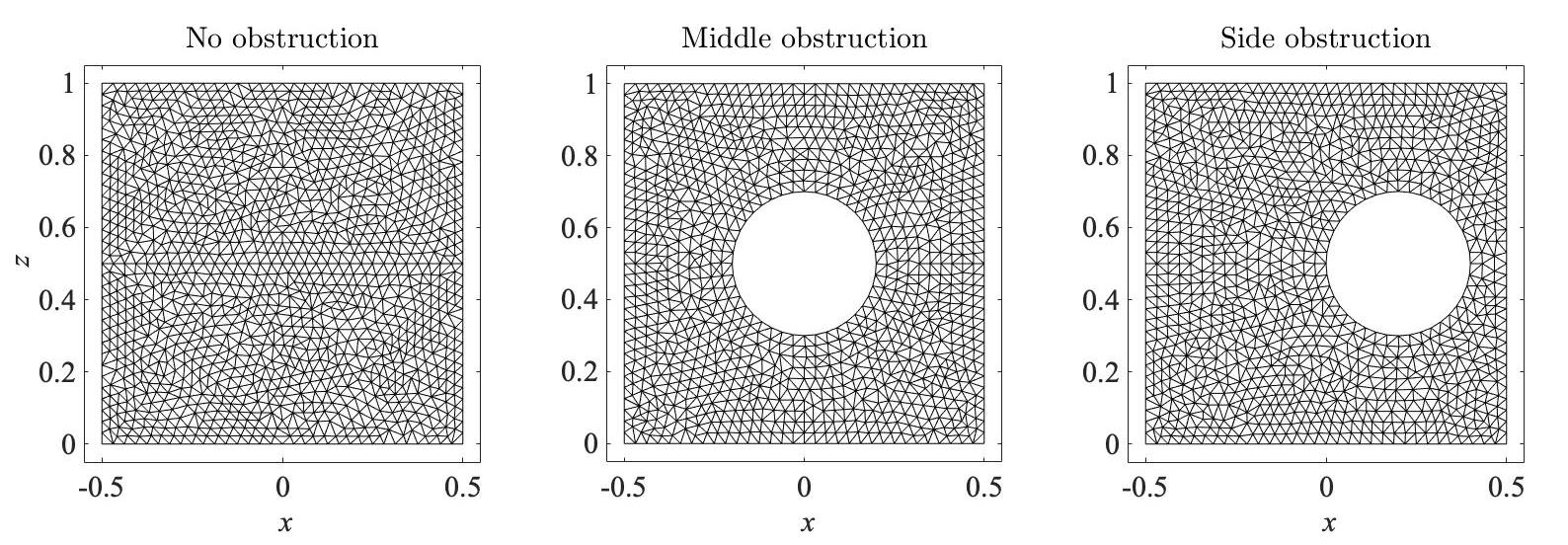}
    \caption{Schematic of meshes for domains with no obstruction, middle obstruction, and side obstruction, respectively.}
    \label{fig:mesh}
\end{figure}
\newline

The weak form of the system of PDEs is evaluated and satisfied on all elements of the domain. We obtain the weak forms by multiplying the Navier-Stokes-Boussinesq equation by a test function $\vv$ and by integrating over the domain $\Omega$. We also add in the incompressibility condition multiplied by the scalar test function for pressure, $q$: 

\begin{multline*}
\int_{\Omega}\frac{\partial \vu}{\partial t}\cdot \vv\,d\Omega +\int_{\Omega}\left(\vu\cdot\nabla\right)\vu\,\cdot \vv\,d\Omega +\int_{\Omega}\nabla p \cdot \vv \,d\Omega -\int_{\Omega}\left(\frac{\Pra}{\Ra}\right)^{\frac{1}{2}} \nabla^2 \vu \cdot \vv\,d\Omega -\int_{\Omega}\theta\, \vk \cdot \vv\,d\Omega \\
+\int_{\Omega}q \nabla\cdot \vu \,d\Omega  = 0\,.
\end{multline*} 

The same procedure is applied to get the weak form of the advection-diffusion equation for heat, where we multiply the equation by a test function $\psi$ and integrate over the domain $\Omega$ to get:
\begin{align*}
    \int_{\Omega}\frac{\partial \theta}{\partial t}\, \psi\,d\Omega + \int_{\Omega}\vu\cdot\nabla \theta\, \psi\,d\Omega -\int_{\Omega}\frac{1}{(\Ra\Pra)^{\frac{1}{2}}} \nabla^2 \theta\, \psi\,d\Omega = 0\,.
\end{align*}

To approximate solutions to the weak forms, we use the method outlined in \cite{mccurdy2020convection} and we introduce the following finite element (FE) spaces:
\begin{itemize}
    \item $V = \{ \vv \in \left[H^1\left(\Omega \right)\right]^2: \vv=\mathbf{0} \textrm{ along the boundaries of }\Omega \}$,\\
    \item $Q = \{ q \in L^2\left(\Omega \right): \int_{\Omega} q\,d\textbf{x}=0 \} =L^2_0\left( \Omega\right)$,\\
    \item $\Psi = \{ \psi \in H^1\left(\Omega \right): \psi =1 \textrm{ on bottom},  \psi =0 \textrm{ on top} \}$.\\
\end{itemize}

We take the initial condition to the be the conductive state, which we perturb to begin the simulations. The perturbation is a seeded random perturbation field, $\epsilon_{mag} := \epsilon_{mag}(\textbf{x})$, with magnitude of $10^{-8}.$ The $\epsilon_{mag}$ term and its effects are discussed in more depth in the following section. So, the simulations begin with
\begin{align}
\left(\vu^{(0)}, \theta^{(0)}\right) = \left( \mathbf{0} ,\, 0.5-z \right) + \epsilon_{mag}\,.\label{eq:ICnumerical}
\end{align}

Given $\left(\vu^{(n)}, \theta^{(n)}\right)\in V\times \Psi$, we find $\left(\vu^{(n+1)}, p^{(n+1)},\theta^{(n+1)}\right)\in V\times Q\times \Psi$ such that

\begin{align}
&\int_{\Omega}\parD{\vu^{(n+1)}}{t}\cdot \vv\,d\xv + \int_{\Omega}\left( \vu^{(n)}\cdot \nabla \right)\vu^{(n+1)}\cdot \vv\,d\xv + 2\left(\frac{\Pra}{\Ra}\right)^{\frac{1}{2}}\int_{\Omega} \mathbb{D}\left(\vu^{(n+1)} \right):\mathbb{D}\left(\vv\right)\,d\xv \nonumber \\[10pt]
    &-\int_{\Omega}\left(\nabla \cdot \vu^{(n+1)}\right)\cdot q\,d\xv  -\int_{\Omega}\left(\nabla \cdot \vv \right)\cdot p^{(n+1)}\,d\xv -\int_{\Omega}\theta^{(n)}\textbf{k}\cdot \vv\,d\xv = 0 \label{eq:Fluid1} \\ \nonumber
\end{align} 

\noindent for all test functions $\vv \in V$ and $q\in Q$, where we have the tensor $\mathbb{D}(\vu) = \frac{1}{2}\left( \nabla \vu + \nabla \vu^{\,\intercal}\right)$, and 

\begin{align}
    \int_{\Omega}\parD{\theta^{(n+1)}}{t} \, \psi\,d\xv+ \int_{\Omega}\vu^{(n+1)}\cdot\nabla \theta^{(n+1)} \, \psi\,d\xv  + \int_{\Omega} \nabla \theta^{(n+1)}\cdot \nabla \psi\,d\textbf{x} = 0 \label{eq:Fluid2}\\ \nonumber
\end{align} 

\noindent for all test functions $\psi \in \Psi$. Although the Navier-Stokes-Boussinesq is nonlinear, we are solving a linear problem in \eqref{eq:Fluid1} since we use the previous velocity $\vu^{(n)}$ in the convective term to help find $\vu^{(n+1)}$. Additionally, the introduction of the $\mathbb{D}(\vu)$ tensors result in a symmetric problem. Since the problem is then linear and symmetric, we can use more computationally efficient solvers to approximate solutions. We also time-lag the nonlinear term of the ADE (since $\vu^{(n+1)}$ will already be known), and thus, we have a decoupled system to solve numerically. For efficiently handling the convective terms in our system, we use a Characteristic Galerkin method (see \cite{glowinski1992finite}) implemented with FreeFem \cite{freefemCite}.

Additionally, we can calculate the stream function $\phi$, by solving 
\begin{align}
    \int_{\Omega} \left[ \nabla\phi^{(n+1)} \cdot \nabla \varphi - \varphi\left(\nabla \times \vu^{(n+1)} \right)\right] d\textbf{x} = 0 \label{eq:Fluid3}\\ \nonumber
\end{align} 

\noindent for all test functions $\varphi \in \Phi$, with the FE space:
\begin{itemize}
\item $\Phi=\{\varphi \in H^1\left(\Omega \right): \varphi=0 \textrm{ along the boundaries of }\Omega \}$.\\
\end{itemize}

To determine when the flow has achieved a steady-state, we calculate the difference between successive velocities and temperatures, where the superscript denotes the time-step of the solution approximations, and require it be less than a specified tolerance of $10^{-6}$. That is, we require
\begin{equation}
   \sqrt {\sum_{i,j} \lvert \vu_{i,j}^{(n+1)} - \vu_{i,j}^{(n)}\rvert^2 + \sum_{i,j} \lvert \theta_{i,j}^{(n+1)} - \theta_{i,j}^{(n)}\rvert^2 } < 10^{-6} \,, \label{eqn:steadyCheck}
\end{equation}
where $n$ is the iteration number and $(i,j)$ are the spatial coordinates. For a given Rayleigh number, once the system achieves a steady-state, the local Nusselt numbers along the $z=0$ boundary are evaluated. All of our simulations are run to a steady-state, or have a final time specified at the beginning of each run. With all of the simulations, the Prandtl number is kept constant at $\Pra = 0.71$, and three different Rayleigh numbers (Ra = $10^4,\, 10^5,\,10^6$) are used to carry out our computations.

Our algorithm is detailed in Algorithm \ref{alg:solvingSystem}. All the time derivatives are discretized with Forward Euler, although a higher-order time integrator could be substituted if desired.

\begin{algorithm}
\caption{Solving the Navier-Stokes-Boussinesq equations coupled with the ADE }
\label{alg:solvingSystem}
\begin{algorithmic}
\STATE{Use initial conditions: $\left(\vu^{(0)}, \theta^{(0)}\right) = \left( \mathbf{0},\, .5-z\right) +\epsilon_{mag}$
.}
\FOR{$n=0;\,n<N\,;n++$}
\STATE{With $\vu^{(n)},\,\theta^{(n)}$, solve \eqref{eq:Fluid1} for $\vu^{(n+1)},\,p^{(n+1)}$.}
  \STATE{With $\vu^{(n+1)},\,\theta^{(n)}$, solve \eqref{eq:Fluid2} for $\theta^{(n+1)}$.}
  \STATE{With $\vu^{(n+1)}$, solve \eqref{eq:Fluid3} for $\phi^{(n+1)}$.}
  \STATE{Check steady-state condition \eqref{eqn:steadyCheck}. {\bf BREAK} if satisfied.}
  \ENDFOR
\RETURN $\vu^{(N)}, \,p^{(N)},\,\theta^{(N)}, \,\phi^{(N)}\,.$
\end{algorithmic}
\end{algorithm}

\section{Code validation}
\begin{figure}[!h]
    \centering
    \includegraphics[height=2.75in]{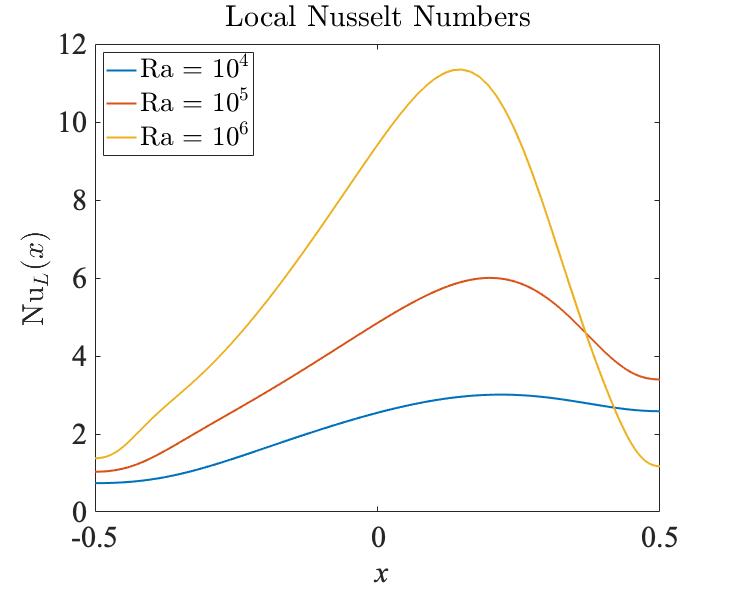}
    \caption{Local Nusselt number on the $z=0$ boundary.}
    \label{fig:nuseltNoObstruction}
\end{figure}

The validity of our numerical method is tested using a square domain without an obstruction, and we compare our results to the benchmark solutions from \cite{ouertatani2008numerical}.
Figure \ref{fig:nuseltNoObstruction} details the local Nusselt numbers for our simulations, and we note a good agreement, visually at least, between our work and \cite{ouertatani2008numerical}. Both graphs present the same trends: the maxima of the local Nusselt numbers increase as the Rayleigh number increases with a shift toward the right of the domain.

\begin{table}[!h]
\centering
\def\arraystretch{1.25}
\begin{tabular}{|l||l|c|c|c|}
\hline
                                           & References                               & $10^4$       & $10^4$      & $10^4$       \\ [0.5ex] 
 \hline\hline
\multirow{2}{*}{Location of $\max\Nus_L$} & Present work                             & $\tilde{x}=0.72$ & $\tilde{x}=0.7$ & $\tilde{x}=0.64$ \\ \cline{2-5} 
                                           & Ouertatani {\it et al.}, \cite{ouertatani2008numerical} & $x=0.7183$       & $x=0.6993$      & $x=0.6448$       \\ [0.5ex] 
 \hline\hline
\multirow{2}{*}{Value of $\max\Nus_L$} & Present work                             & $3.014$          & $6.015$         & $11.50$          \\ \cline{2-5} 
                                           & Ouertatani {\it et al.}, \cite{ouertatani2008numerical} & $3.023$          & $6.065$         & $11.69$ \\ \hline
\end{tabular}
\caption{\label{tab:ValidationResults} The location and value of the maxima of the local Nusselt numbers for validation. In the present work, we note the $x-$location with the spatial offset of $\tilde{x}=x+0.5$ to align our data with \cite{ouertatani2008numerical}.}
\end{table}

\vspace{-.1in}

Table \ref{tab:ValidationResults} presents the location and values of of the maximum local Nusselt numbers from the present study and the work of \cite{ouertatani2008numerical}. In comparing the location of the maxima between our work and \cite{ouertatani2008numerical}, we note that since our domain is from $-0.5 \leq x \leq 0.5$ and the domain considered by Ouertatani {\it et al.} is $0 \leq x \leq 1$, we offset our data points by $\tilde{x}=x+0.5$ to align the data points. Once again, we observe good agreement with small deviations in our results, likely due to the selection of mesh points.

The perturbation $\epsilon_{mag}$ of \eqref{eq:ICnumerical} is what allows the unstable conductive state to evolve into the stable convective state. So that we are able to compare our results to \cite{ouertatani2008numerical}, we used a seeded random perturbation that allowed us to recover the same direction of flow as the benchmarks by Ouertatani {\it et al.}. The same perturbation was used in remaining simulations of this work as well. If the perturbation broke symmetry in a way where the flow circulated in the opposite direction, the local Nusselt values of Figure \ref{fig:nuseltNoObstruction} would appear symmetric about $x=0$.

\begin{figure}[!h]
    \centering
    \includegraphics[width=\textwidth]{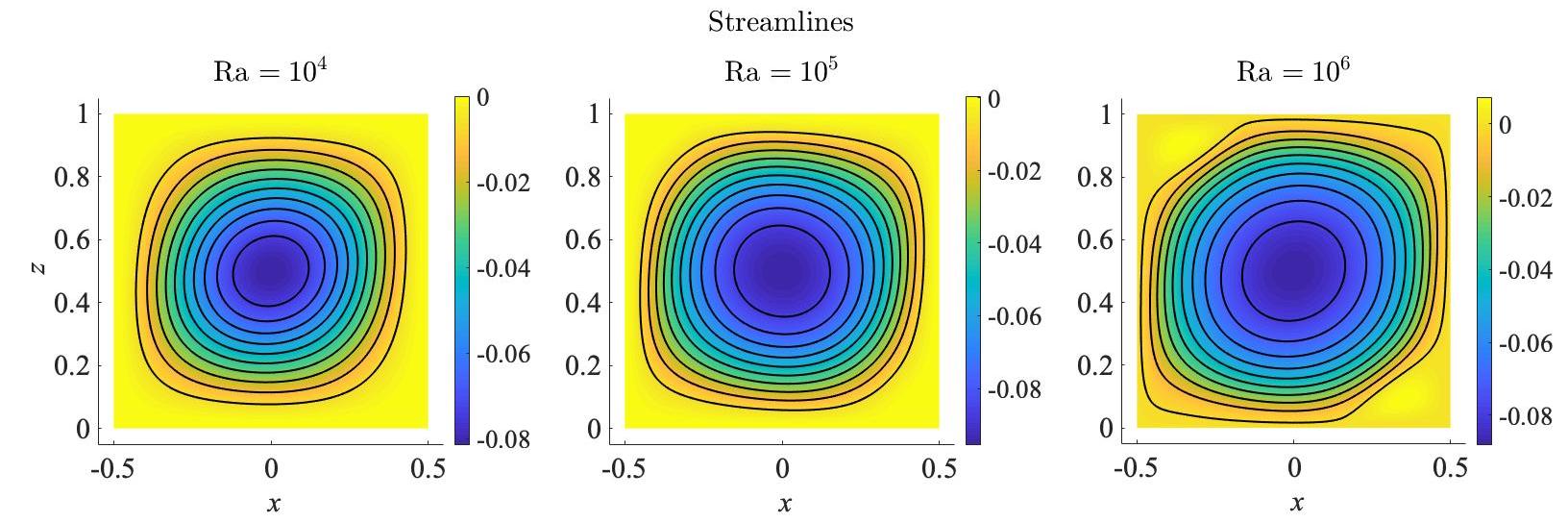}
     \includegraphics[width=\textwidth]{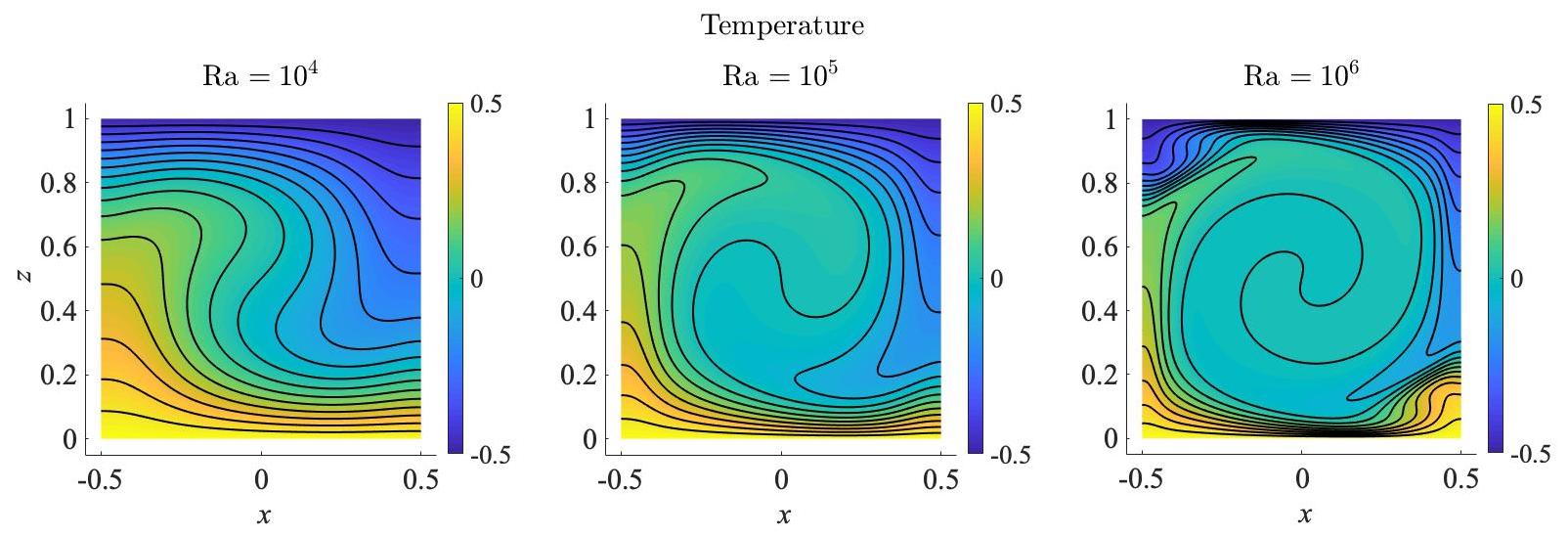}
         \includegraphics[width=\textwidth]{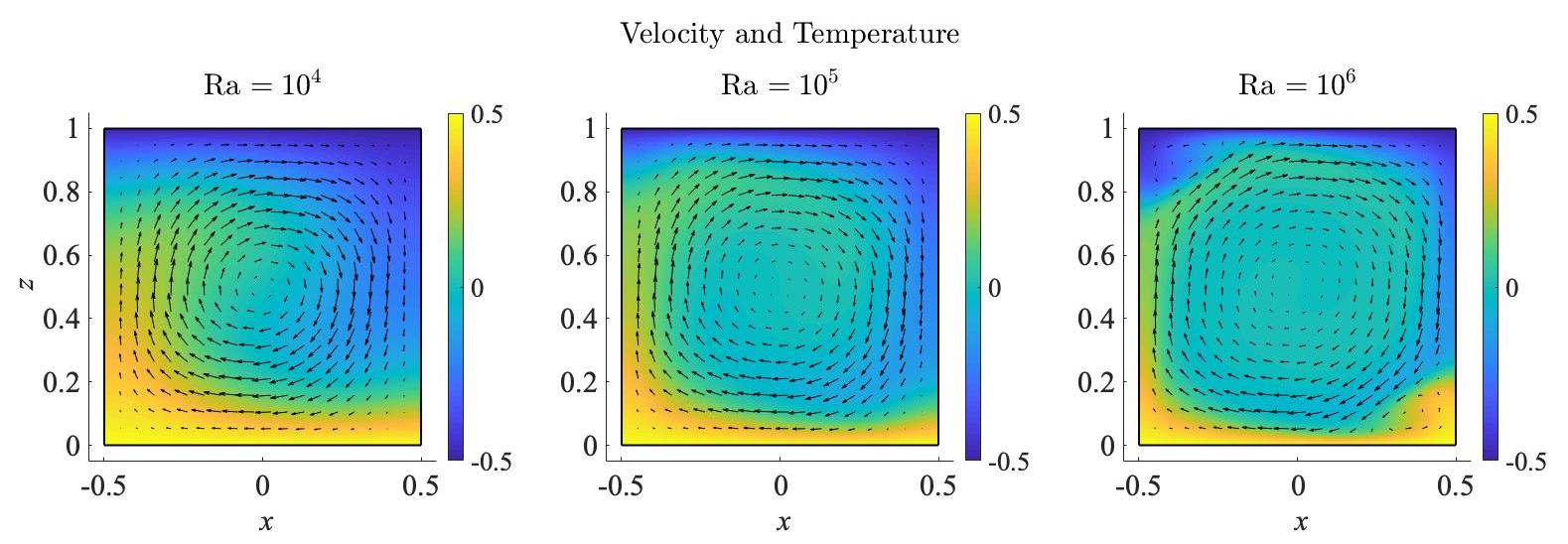}
    \caption{Top row: streamline contours, middle row: temperature profile contours, bottom row: velocity field and temperature fields. Each column represents a different Rayleigh number, with $\Ra = 10^4, \Ra = 10^5,$ and $\Ra=10^6$ as the three columns, respectively.}
    \label{fig:NoObstruction}
\end{figure}

In addition to quantitative features of the flows like the local Nusselt values, we also compare qualitative features: streamline contours, temperature profiles, and velocity fields. Streamlines are curves that are tangent to the flow's velocity vector and depict the path that the fluid takes at any given time. The top row of graphs in Figure \ref{fig:NoObstruction} shows the streamline contours for three different Rayleigh numbers considered. The negative value associated with the contours notes that the fluid is moving in the clockwise direction with the magnitude noting the speed of circulation. The first two cases with $\Ra=10^4$ and $\Ra=10^5$ show uniform clockwise movement, while the last case with $\Ra=10^6$ exhibits a small amount of counter-clockwise rotation in the upper left and bottom right corners of the domain, as noted by the slightly positive values of the streamline function there.

The second row of Figure \ref{fig:NoObstruction} presents the temperature profiles of the same simulations, and we observe clockwise circular motion in agreement with the streamline contours in the top row. We see the cooler fluid from the top surface sinking along the right of the domain while the hotter fluid from the bottom surface rises up along left boundary. As the Rayleigh number increases and the convection cells circulate faster, the temperature deviations from the linear conductive state become more pronounced as well.

In the last row of Figure \ref{fig:NoObstruction}, we show the velocity and temperature fields. The length of the arrows is proportional to the magnitude of the velocity and the direction depicts the direction in which the fluid moves. Once again, the clockwise movement observed agrees with the streamline and temperature profiles. 

With the similar profiles found by Ouertatani {\it et al.} \cite{ouertatani2008numerical}, we are confident that our numerical methods are working as they should in approximating solutions. Next, we move on to simulations with an obstruction present in the flow, although we will still use the three qualitative features (streamlines, temperature, velocity fields) to help us categorize and measure convection. Additionally, we will use three quantitative markers: the local Nusselt number (for comparison to \cite{ouertatani2008numerical}), and the Nusselt number and mathematical energy.

\section{Results and discussions}
To answer our research question on whether an obstruction in an enclosure affects the convection in the fluid, there are several factors to be considered. These factors include the size of the obstruction, its shape, its location within the enclosure, and the number of obstructions, as well as their relative positions to each other within the fluid. These factors should all be evaluated to fully answer the question. For this paper we investigate how the location of a circular obstruction affect convection. However, for future works we will investigate other aspects that affect convection such as size of the obstruction or an optimal location to hinder, or enhance, convection.

\begin{figure}[]
    \centering
    \includegraphics[width=\textwidth]{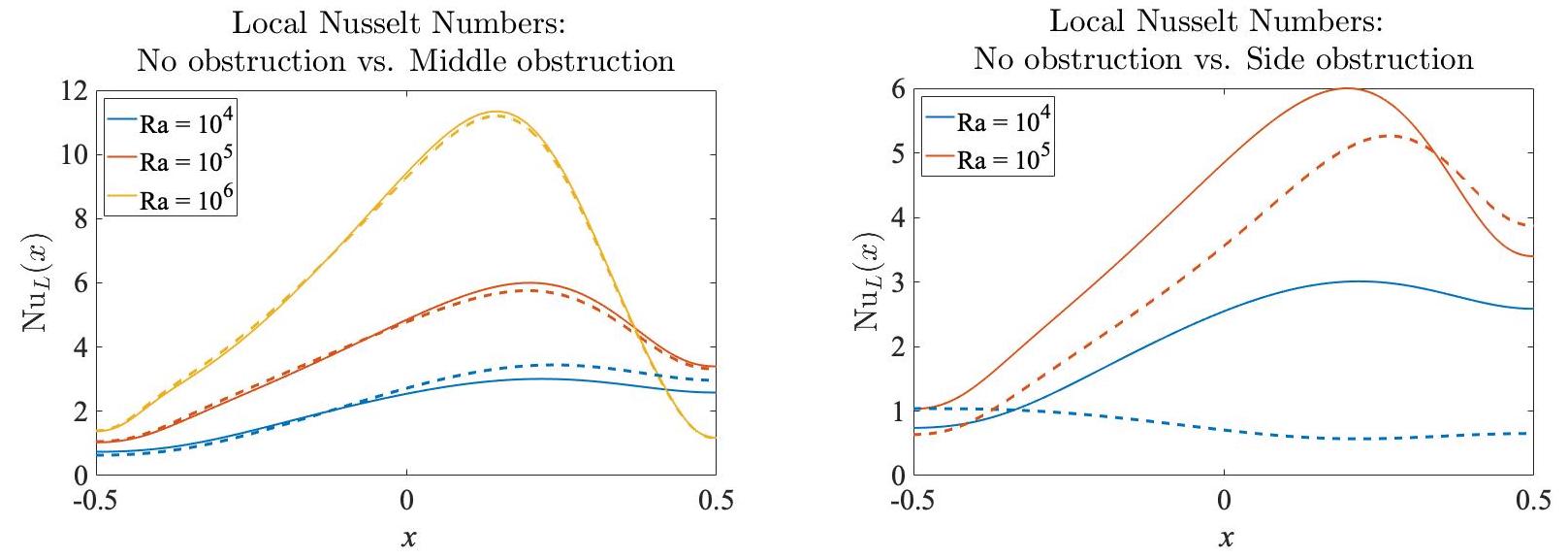}
    \caption{Comparing local Nusselt number for the cases with no obstruction (solid lines) and the middle obstruction and side obstruction (dashed lines) for the first and second panels, respectively. The local Nusselt values are calculated at steady-states, so no profile is calculated for the side obstruction at $\Ra=10^6$ since it does not achieve a steady-state.}
    \label{fig:localNusselt}
\end{figure}

When the obstruction is placed in the center of the domain, we observe that the nature of convection in the fluid is not greatly affected. As shown in the first graph of Figure \ref{fig:localNusselt}, the curves for the local Nusselt numbers along the lower boundary for the cases with the middle obstruction behave in a similar way to those with no obstruction. Thus, we conclude that the presence of an obstruction in a flow does not always result in significant differences in flow behavior.

However, when the same obstruction is placed towards the side of the domain (shown in the third schematic of Figure \ref{fig:mesh}), we observe that the effect on the nature of convection differs depending on the temperature difference, which is correlated to the nondimensional constant $\Ra$. We notice that convection for $\Ra = 10^4$ with the side obstruction is qualitatively different from the case without an obstruction. This difference is noted by the variation in the shape of the curves for $\Ra = 10^4$ as seen in the second plot of Figure \ref{fig:localNusselt}, where the no-obstruction case has a maximum around $\Nus_L\approx 3$ and the local Nusselt number hovers around 1 for the side-obstruction case. The profile of $\Nus_L\approx 1$ signifies that there is little convective heat transfer at the lower boundary, and likely no convection cells forming in the system for this parameter regime and obstruction location.

For the cases with $\Ra = 10^5$, convection in the fluid is not affected as much since the no-obstruction and side-obstruction $\Nus_L$ curves behave similarly. For a higher temperature difference of $\Ra = 10^6$, we find the system does not reach a steady-state. Since no steady-state is reached, the local Nusselt values are not calculated and there is no curve for this case presented. Therefore, we turn to the mathematical energy and Nusselt numbers as a function of time to help describe this flow. 

\begin{figure}[!h] 
    \centering
    \includegraphics[width=\textwidth]{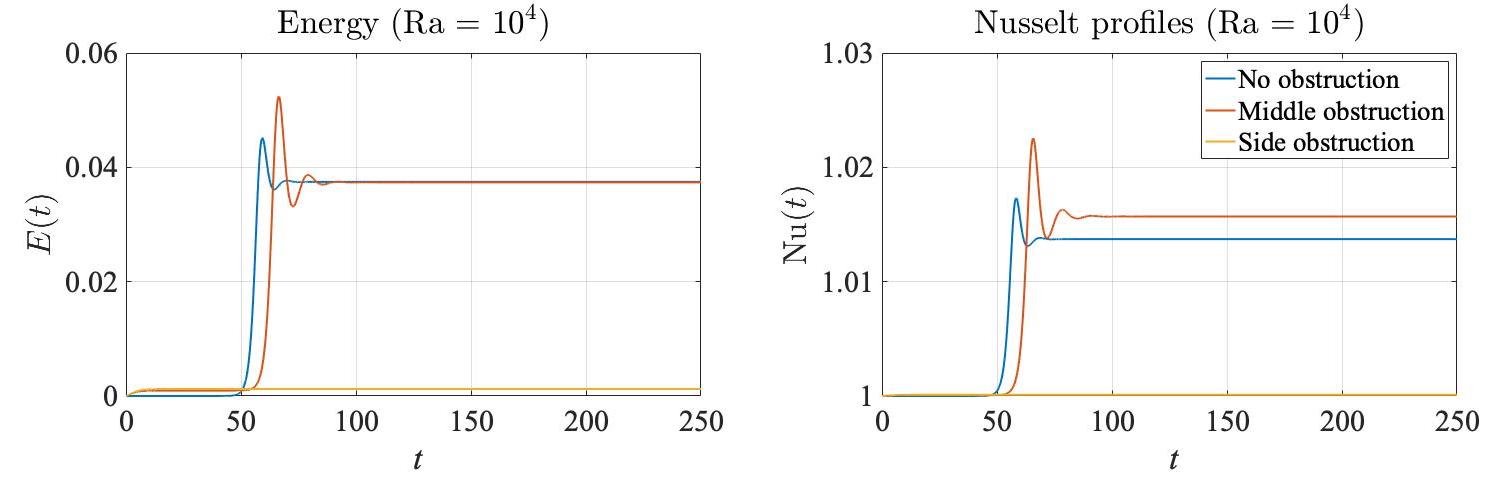}
     \includegraphics[width=\textwidth]{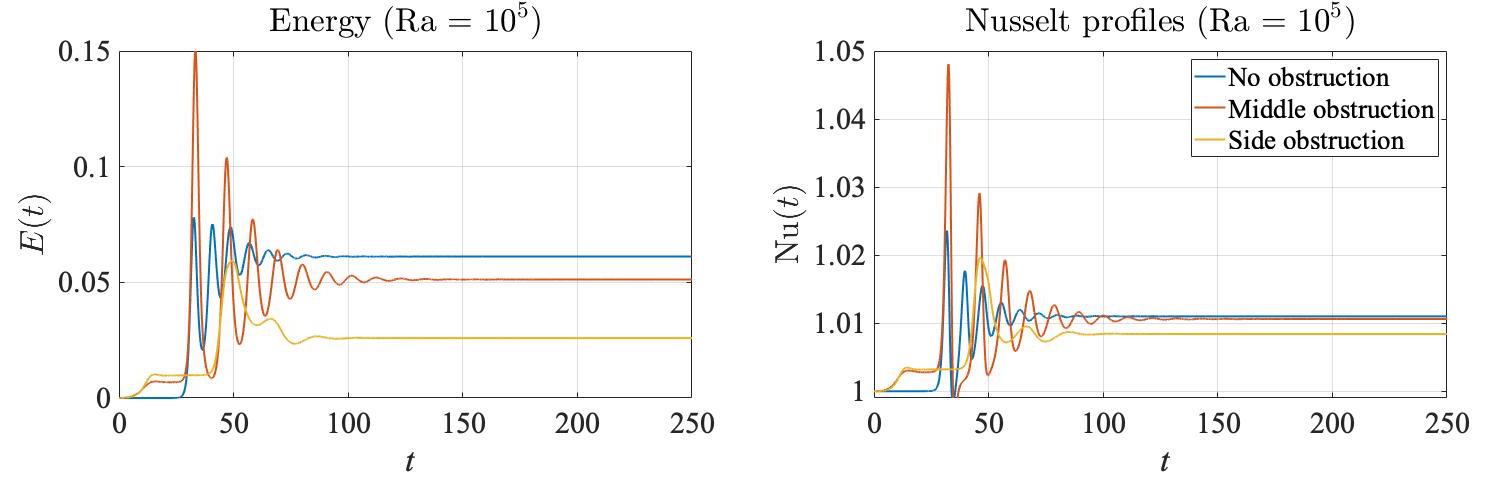}
      \includegraphics[width=\textwidth]{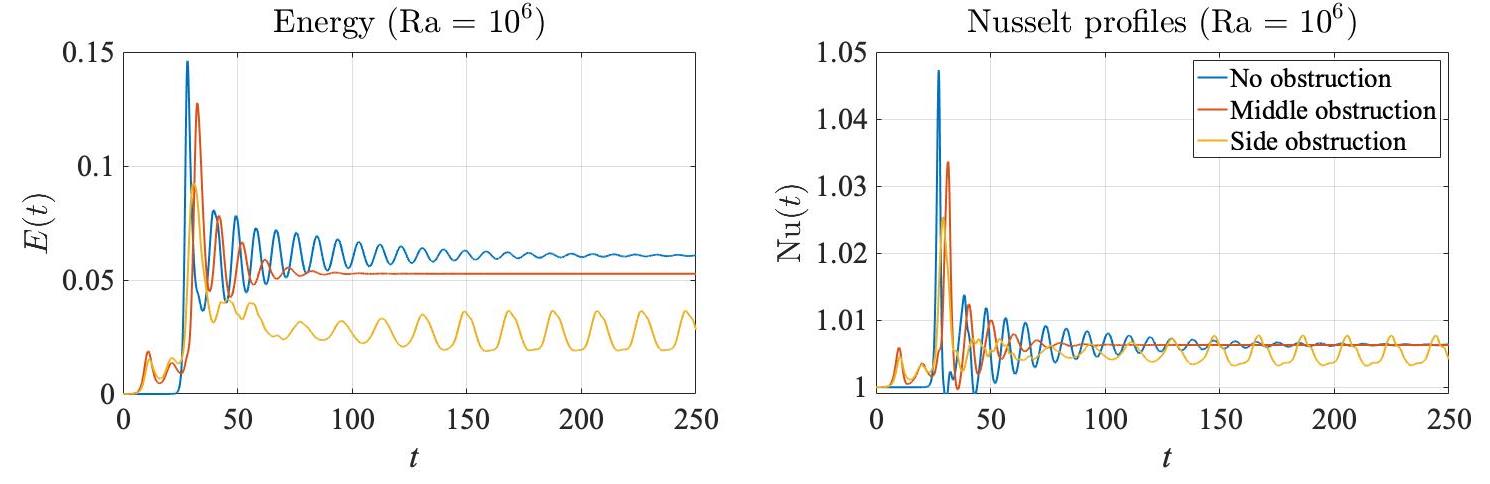}
    \caption{Energy profiles and Nusselt numbers for $\Ra =10^4$ in first row, $\Ra =10^5$ in the second row, and $\Ra =10^6$ in the third row.}
    \label{fig:energyAndNusselt}
\end{figure}

The graphs in Figure \ref{fig:energyAndNusselt} compare the energies and Nusselt numbers for each of our cases, with each row comparing simulations of the same Rayleigh number. For $\Ra = 10^4$ (seen in the top row), the energy trend for the cases without an obstruction and the middle obstruction cases are similar. However for the side obstruction case, there is barely any mathematical energy in the system and the Nusselt number is $\Nus(t)=1$ for the duration of the simulation. While the temperature field deviates slightly from the linear conductive state (since it must satisfy the Neumann conditions around the boundary of the obstruction), there is  little fluid velocity. The temperature and vertical fluid velocity are both symmetric about $z=.5$(evidenced by the patterns of the temperature profiles and streamlines for this case in Figure \ref{fig:SteadyCases}), and this symmetry allows the Nusselt number calculated by Equation \eqref{eqn:nuss} to be close to 1, signifying there is no convective heat transfer. Therefore, the location of the side obstruction hinders convection almost entirely, which agrees with the results from the local Nusselt profiles shown in Figure \ref{fig:localNusselt}. 

For $\Ra = 10^5$, as seen in the second row of Figure \ref{fig:energyAndNusselt}, the trends in the mathematical energies and Nusselt numbers are relatively similar for all three cases in that they first oscillate and then settle down to a steady-state. 

With $\Ra = 10^6$ in the last row of Figure \ref{fig:energyAndNusselt}, we find the side obstruction case exhibits qualitatively different flow behavior than the other two cases. We see the energy for the no-obstruction and middle obstruction cases level out in time as the systems reach their steady states. However, the side obstruction case does not achieve a steady-state for this Rayleigh number; it exhibits periodic, oscillatory behavior. This unsteady, periodic flow for $\Ra=10^6$ is reminiscent of the Von K\'{a}rm\'{a}n vortex shedding in the case of considering flow past a cylinder; for flows with a Reynolds number $\textrm{Re}< 47,$ flow past the cylinder is steady, but for flows with $\textrm{Re}>47,$ unsteady, periodic eddies begin to form in the wake of the cylinder, as explained in \cite{strykowski1990formation}. In the Appendix, we show the temperature and streamline profiles at different intervals of one period of the unsteady behavior. 

One interesting takeaway from our investigation deals with the similarities in the energy of the system and the Nusselt profiles shown in Figure \ref{fig:energyAndNusselt}. Traditionally, the Nusselt number is used to convey how much convective heat transfer is present in a system. However, we see that the mathematical energy mimics the behavior of the physically-motivated Nusselt number, and can also be used in linear and nonlinear stability analyses to determine the critical Rayleigh number for flows in various settings. Therefore, the mathematical energy of the system could be a more flexible marker of convection than the Nusselt number. A future direction of this project could try to show that $E(t)$ and $\Nus(t)$ are equivalent in some sense, likely obtained with some scaling argument.

\section{Conclusion, and future work}
In this work, we used a finite element method to investigate how the location of an obstruction affect convection in a fluid enclosed in a square domain. To begin our numerical investigation and validate the accuracy of our code, we first considered natural convection inside a square {\it without} an obstruction and compared our results those from \cite{ouertatani2008numerical}. After establishing that our numerical methods worked as expected, we investigated how the location of an obstruction affected flow profiles. To answer this question, we ran simulations for a flows without an obstruction and compared them to two cases with an obstruction-- one with an obstruction at the center of the domain, and another with the obstruction's placement towards the side (as seen in Figure \ref{fig:mesh}). 

We compared results from our simulation for the different three cases of no obstruction, middle obstruction and side obstruction. From a quantitative view, we plotted graphs of local Nusselt numbers on the lower boundary for the Rayleigh numbers $\Ra =10^4$, $\Ra =10^5$, and $\Ra =10^6$ respectively, comparing the curves for middle obstruction and side obstruction to the no-obstruction case as shown in Figure \ref{fig:localNusselt}. We also compared mathematical energy profiles and Nusselt numbers in Figure \ref{fig:energyAndNusselt} to analyze flow behavior as it evolved in time. 

With the results of these quantitative markers, we find that placing an obstruction at the center of the square domain does not greatly affect convection. However, when the same obstruction was moved to the side of the domain, the effect on convection depended on the temperature difference in the system. Notable examples showed that the side obstruction could prohibit the onset of convection for $\Ra=10^4$ and resulted in periodic, unsteady flow for $\Ra=10^6$-- both of which were drastically different than their corresponding simulations with no obstruction or the middle-obstruction cases.

Future work will investigate how the size and location of a circular obstruction affects the convection in the fluid enclosed in a square domain. Additionally, this study can further be expanded to researching nature of natural convection in a fluid with an obstruction for $3$D cases, or studied with different boundary conditions on the obstruction or along the borders of the enclosure. 

In the Appendix, we present different results with streamlines, temperature profiles, and velocity fields for the various cases as benchmarks for possible comparison in future research.

\section*{Acknowledgments}
We would like to acknowledge the Trinity College Summer Research Program for funding.

\bibliographystyle{siamplain}
\bibliography{bibliography}

\begin{thebibliography}{10}

\bibitem{adachi1996stability}
{\sc T.~Adachi and J.~Mizushima}, {\em Stability of the thermal convection in a
  tilted square cavity}, Journal of the Physical Society of Japan, 65 (1996),
  pp.~1686--1698.

\bibitem{benard1900etude}
{\sc H.~B{\'e}nard}, {\em Etude exp{\'e}rimentale du mouvement des liquides
  propageant de la chaleur par convection. {R}{\'e}gime permanent: tourbillons
  cellulaires}, Compte-rendus de l’Acad{\'e}mie des sciences, 130 (1900),
  pp.~1004--1007.

\bibitem{benard1900mouvements}
{\sc H.~B{\'e}nard}, {\em Mouvements tourbillonnaires {\`a} structure
  cellulaire. {\'e}tude optique de la surface}, Compte-rendus de
  l’Acad{\'e}mie des Sciences, 130 (1900), pp.~1065--1068.

\bibitem{benard1900tourbillons}
{\sc H.~B{\'e}nard}, {\em Tourbillons c{\'e}llularies dans une nappe liquide:
  Pt. {I}, {D}escription g{\'e}n{\`e}rale des ph{\'e}nom{\`e}nes; {P}t. {II},
  {P}recedes mecaniques et optiques d’examen; lois numeriques des
  phenomenes}, Rev. Gen. Sci. Pure Appl, 11 (1900), pp.~1261--1309.

\bibitem{benard1901tourbillons}
{\sc H.~B{\'e}nard}, {\em Les tourbillons cellulaires dans une nappe liquide
  propageant de la chaleur par convection: en r{\'e}gime permanent},
  Gauthier-Villars, 1901.

\bibitem{benard1901paper}
{\sc H.~B{\'e}nard}, {\em Les tourbillons c{\'e}llularies dans une nappe
  liquide propageant de la chaleur par convection en r{\'e}gime permanent},
  Annales de Chimie et de Physique,  (1901).

\bibitem{chandrasekhar1981hydrodynamic}
{\sc S.~Chandrasekhar}, {\em Hydrodynamic and Hydromagnetic Stability}, Dover
  Books on Physics Series, Dover Publications, 1981.

\bibitem{doering1996variational}
{\sc C.~R. Doering and P.~Constantin}, {\em Variational bounds on energy
  dissipation in incompressible flows. {I}{I}{I}. {C}onvection}, Physical
  {R}eview {E}, 53 (1996), p.~5957,
  \url{https://doi.org/10.1103/PhysRevE.53.5957}.

\bibitem{doering2001upper}
{\sc C.~R. Doering and P.~Constantin}, {\em On upper bounds for infinite
  prandtl number convection with or without rotation}, Journal of Mathematical
  Physics, 42 (2001), pp.~784--795, \url{https://doi.org/10.1063/1.1336157}.

\bibitem{getling1998rayleigh}
{\sc A.~V. Getling}, {\em Rayleigh-B{\'e}nard convection: structures and
  dynamics}, vol.~11, World Scientific, 1998.

\bibitem{ghaddar1992natural}
{\sc N.~K. Ghaddar}, {\em Natural convection heat transfer between a uniformly
  heated cylindrical element and its rectangular enclosure}, International
  journal of heat and mass transfer, 35 (1992), pp.~2327--2334.

\bibitem{glowinski1992finite}
{\sc R.~Glowinski and O.~Pironneau}, {\em Finite element methods for
  {N}avier-{S}tokes equations}, Annual {R}eview of {F}luid {M}echanics, 24
  (1992), pp.~167--204,
  \url{https://doi.org/10.1146/annurev.fl.24.010192.001123}.

\bibitem{freefemCite}
{\sc F.~Hecht}, {\em New development in {F}ree{F}em++}, J. Numer. Math., 20
  (2012), pp.~251--265, \url{https://freefem.org/}.

\bibitem{howard1963heat}
{\sc L.~N. Howard}, {\em Heat transport by turbulent convection}, Journal of
  {F}luid {M}echanics, 17 (1963), pp.~405--432,
  \url{https://doi.org/10.1017/S0022112063001427}.

\bibitem{howard1972bounds}
{\sc L.~N. Howard}, {\em Bounds on flow quantities}, Annual Review of Fluid
  Mechanics, 4 (1972), pp.~473--494,
  \url{https://doi.org/10.1146/annurev.fl.04.010172.002353}.

\bibitem{kim2008numerical}
{\sc B.~Kim, D.~Lee, M.~Ha, and H.~Yoon}, {\em A numerical study of natural
  convection in a square enclosure with a circular cylinder at different
  vertical locations}, International journal of heat and mass transfer, 51
  (2008), pp.~1888--1906.

\bibitem{koschmieder1993benard}
{\sc E.~L. Koschmieder}, {\em B{\'e}nard cells and {T}aylor vortices},
  Cambridge University Press, 1993.

\bibitem{lappa2009thermal}
{\sc M.~Lappa}, {\em Thermal convection: patterns, evolution and stability},
  John Wiley \& Sons, 2009.

\bibitem{lee2010natural}
{\sc J.~Lee, M.~Ha, and H.~Yoon}, {\em Natural convection in a square enclosure
  with a circular cylinder at different horizontal and diagonal locations},
  International Journal of Heat and Mass Transfer, 53 (2010), pp.~5905--5919.

\bibitem{lee2016three}
{\sc S.~H. Lee, Y.~M. Seo, H.~S. Yoon, and M.~Y. Ha}, {\em Three-dimensional
  natural convection around an inner circular cylinder located in a cubic
  enclosure with sinusoidal thermal boundary condition}, International Journal
  of Heat and Mass Transfer, 101 (2016), pp.~807--823.

\bibitem{mccurdy2020convection}
{\sc M.~McCurdy}, {\em Convection in Coupled Fluid-Porous Media Systems: A Tale
  of Two Fluids}, PhD thesis, The Florida State University, 2020.

\bibitem{mccurdy2019convection}
{\sc M.~McCurdy, N.~Moore, and X.~Wang}, {\em Convection in a coupled free
  flow-porous media system}, SIAM Journal on Applied Mathematics, 79 (2019),
  pp.~2313--2339, \url{https://doi.org/10.1137/19M1238095}.

\bibitem{miroshnichenko2018turbulent}
{\sc I.~Miroshnichenko, M.~Sheremet, and A.~J. Chamkha}, {\em Turbulent natural
  convection combined with surface thermal radiation in a square cavity with
  local heater}, International Journal of Numerical Methods for Heat \& Fluid
  Flow,  (2018).

\bibitem{mizushima1995onset}
{\sc J.~Mizushima}, {\em Onset of the thermal convection in a finite
  two-dimensional box}, Journal of the Physical Society of Japan, 64 (1995),
  pp.~2420--2432.

\bibitem{mizushima1995structural}
{\sc J.~Mizushima and T.~Adachi}, {\em Structural stability of the pitchfork
  bifurcation of thermal convection in a rectangular cavity}, Journal of the
  Physical Society of Japan, 64 (1995), pp.~4670--4683.

\bibitem{mizushima1997sequential}
{\sc J.~Mizushima and T.~Adachi}, {\em Sequential transitions of the thermal
  convection in a square cavity}, Journal of the Physical Society of Japan, 66
  (1997), pp.~79--90.

\bibitem{time_periodic_cooling}
{\sc L.~Nasseri, N.~Himrane, D.~Ameziani, B.~Abderrahmane, and R.~Bennacer},
  {\em Time-periodic cooling of {R}ayleigh--{B}{\'e}nard convection}, Fluids, 6
  (2021), p.~87, \url{https://doi.org/10.3390/fluids6020087}.

\bibitem{ouertatani2008numerical}
{\sc N.~Ouertatani, N.~B. Cheikh, B.~B. Beya, and T.~Lili}, {\em Numerical
  simulation of two-dimensional {R}ayleigh--{B}{\'e}nard convection in an
  enclosure}, Comptes Rendus M{\'e}canique, 336 (2008), pp.~464--470,
  \url{https://doi.org/10.1016/j.crme.2008.02.004}.

\bibitem{paroncini2009natural}
{\sc M.~Paroncini and F.~Corvaro}, {\em Natural convection in a square
  enclosure with a hot source}, International journal of thermal sciences, 48
  (2009), pp.~1683--1695.

\bibitem{pellew1940maintained}
{\sc A.~Pellew and R.~V. Southwell}, {\em On maintained convective motion in a
  fluid heated from below}, Proceedings of the Royal Society of London. Series
  A. Mathematical and Physical Sciences, 176 (1940), pp.~312--343,
  \url{https://doi.org/10.1098/rspa.1940.0092}.

\bibitem{rayleigh1916}
{\sc L.~Rayleigh}, {\em On convection currents in a horizontal layer of fluid,
  when the higher temperature is on the under side}, Philosophical Magazine and
  Journal of Science (Sixth Series), 32 (1916), pp.~529--546,
  \url{https://doi.org/10.1080/14786441608635602}.

\bibitem{reid1958some}
{\sc W.~Reid and D.~Harris}, {\em Some further results on the {B}{\'e}nard
  problem}, The Physics of fluids, 1 (1958), pp.~102--110,
  \url{https://doi.org/10.1063/1.1705871}.

\bibitem{strykowski1990formation}
{\sc P.~J. Strykowski and K.~R. Sreenivasan}, {\em On the formation and
  suppression of vortex ‘shedding’ at low {R}eynolds numbers}, Journal of
  Fluid Mechanics, 218 (1990), pp.~71--107,
  \url{https://doi.org/10.1017/S0022112090000933}.

\bibitem{TURAN201283}
{\sc O.~Turan, N.~Chakraborty, and R.~J. Poole}, {\em Laminar
  {R}ayleigh--{B}{\'e}nard convection of yield stress fluids in a square
  enclosure}, Journal of Non-Newtonian Fluid Mechanics, 171-172 (2012),
  pp.~83--96, \url{https://doi.org/10.1016/j.jnnfm.2012.01.006}.

\bibitem{van_der_poel_stevens_lohse_2013}
{\sc E.~P. van~der Poel, R.~J. A.~M. Stevens, and D.~Lohse}, {\em Comparison
  between two- and three-dimensional {R}ayleigh--{B}{\'e}nard convection},
  Journal of Fluid Mechanics, 736 (2013), p.~177–194,
  \url{https://doi.org/10.1017/jfm.2013.488}.

\bibitem{zhao2006conjugate}
{\sc F.-Y. Zhao, G.-F. Tang, and D.~Liu}, {\em Conjugate natural convection in
  enclosures with external and internal heat sources}, International Journal of
  Engineering Science, 44 (2006), pp.~148--165.

\end{thebibliography}

\newpage

\section{Appendix}
\label{sect:appendix}
\begin{figure}[!h]
    \centering
    \includegraphics[width=.9\textwidth]{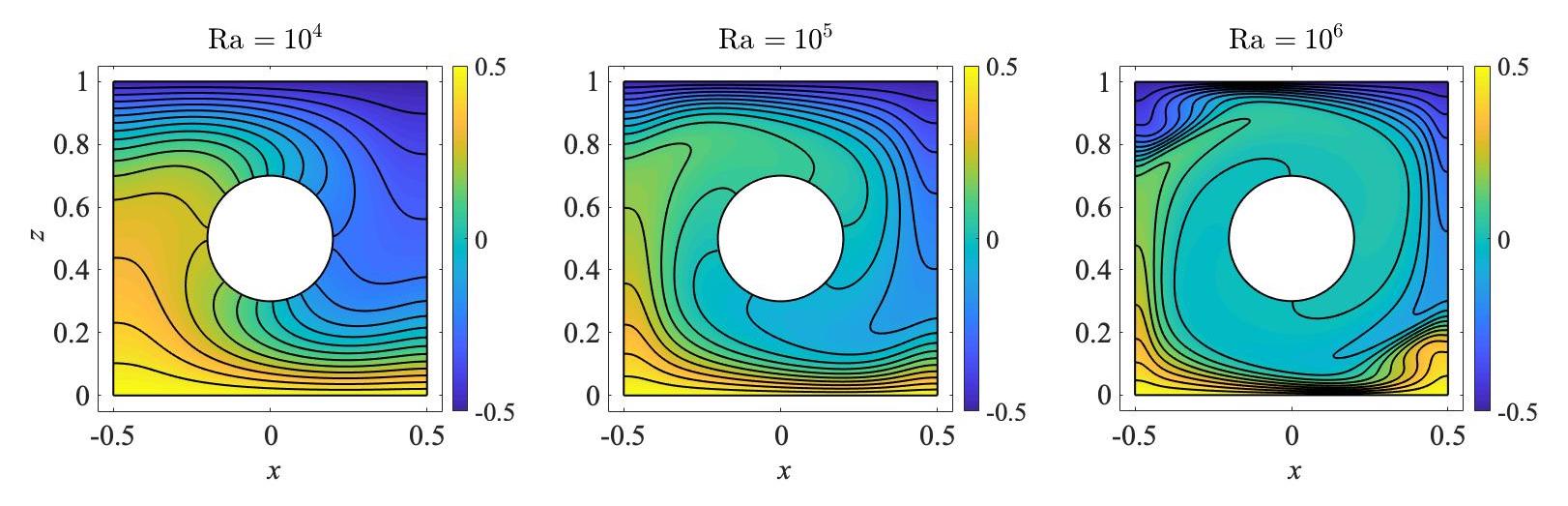}
    \includegraphics[scale=0.24]{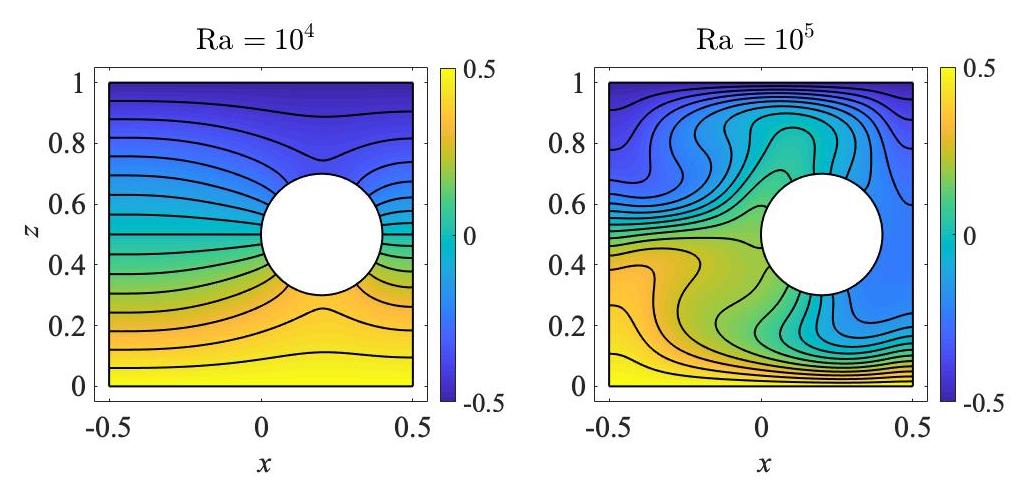}
    \includegraphics[width=.9\textwidth]{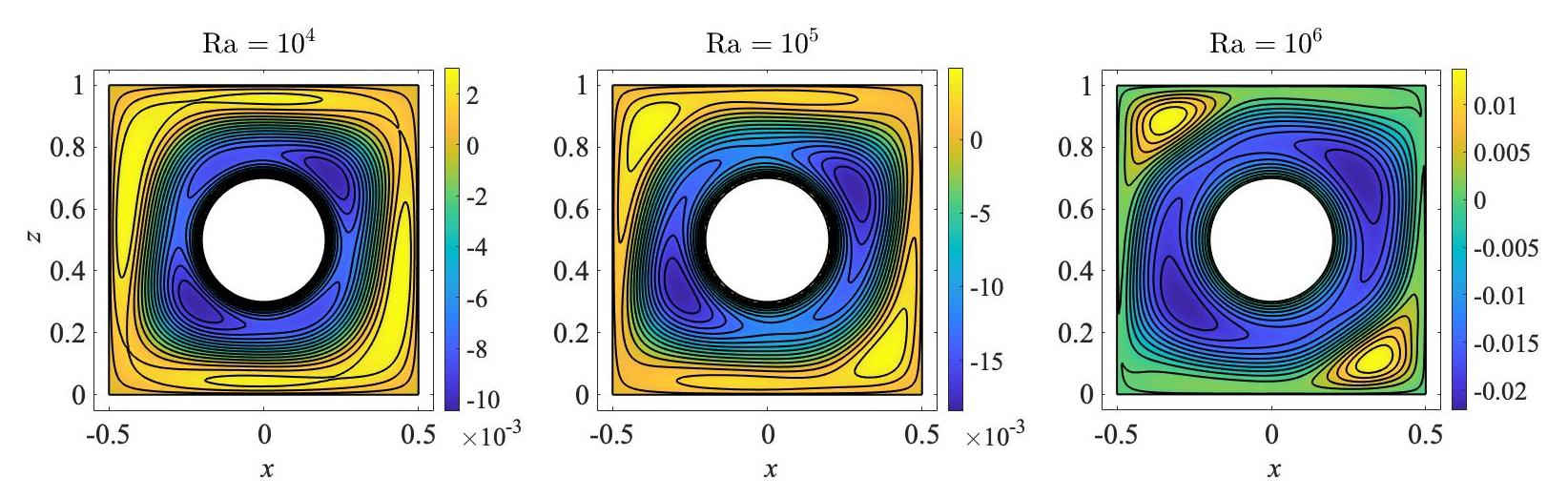}
     \includegraphics[scale=0.24]{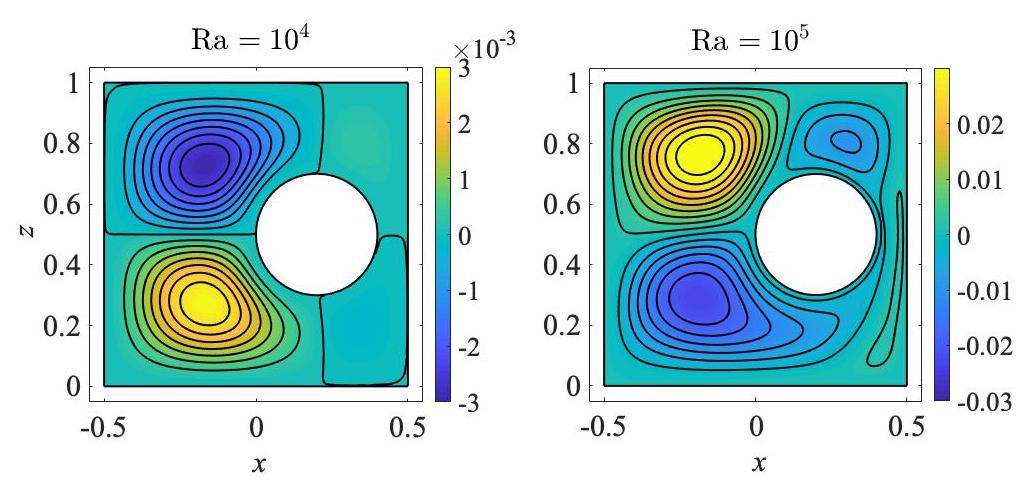}
    \caption{Top two rows: temperature profiles for a middle obstruction with $\Ra=10^4,\,10^5,\,10^6$ at their steady-states, and the side obstruction cases with $\Ra=10^4$ and $\Ra=10^5$ only. Bottom two rows: streamline contours for the same simulations.}
    \label{fig:SteadyCases}
\end{figure}

Other results from our simulations. Figure \ref{fig:SteadyCases} shows the middle and side obstruction cases which reached their steady-states.

\newpage

\begin{figure}[!h]
    \centering
    \includegraphics[scale=0.25]{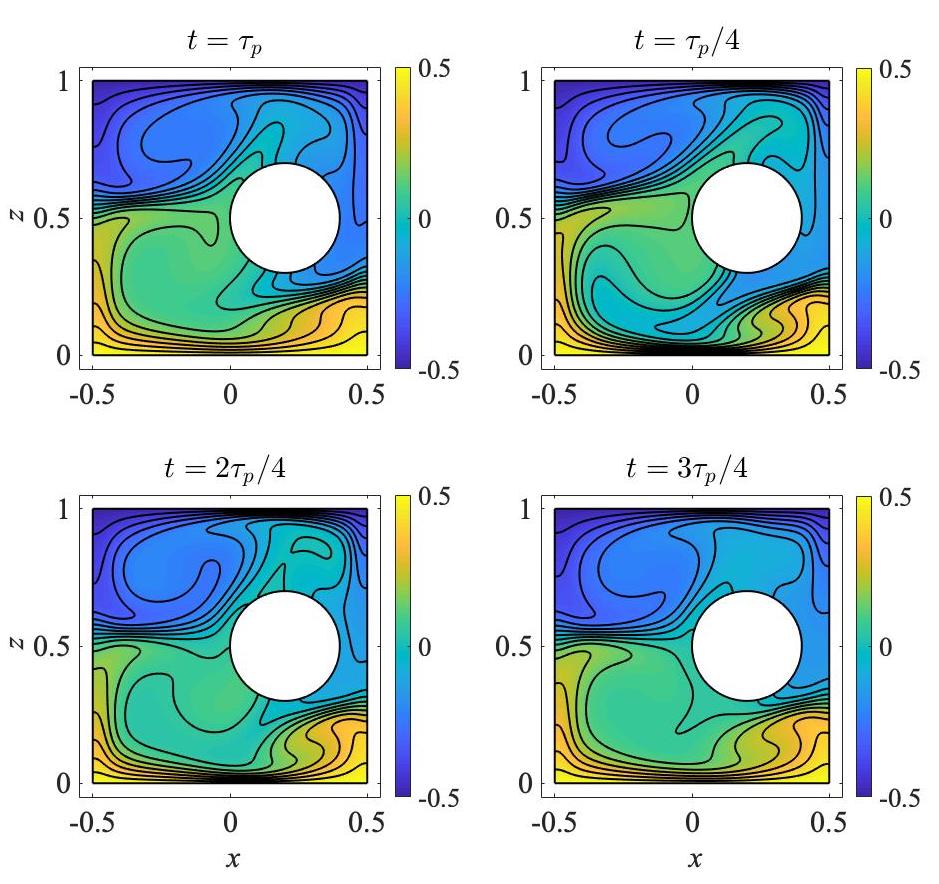}
        \includegraphics[scale=0.25]{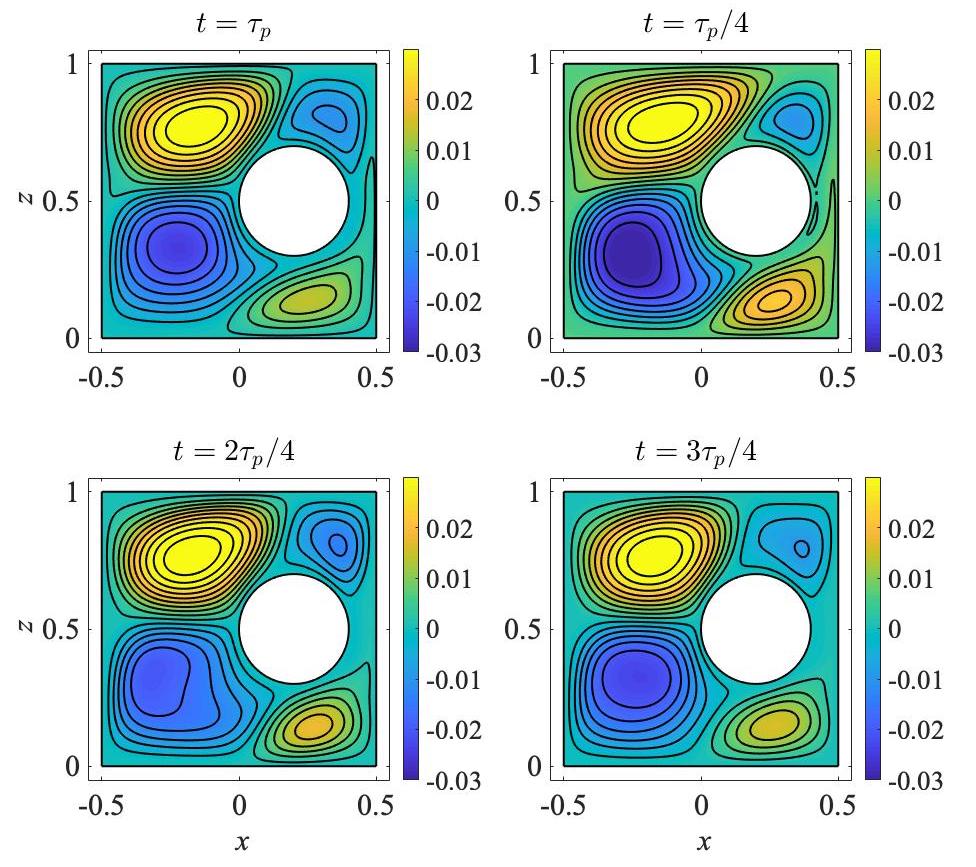}
    \caption{Top two rows: temperatures for side obstruction case with $\Ra = 10^6$ at intervals of one-period in the behavior. Bottom two rows: streamlines for side obstruction case with $\Ra = 10^6$ at the same time steps.}
    \label{fig:UnsteadyCase}
\end{figure}

Figure \ref{fig:UnsteadyCase} shows results from the side obstruction case with $\Ra=10^6$, which exhibited periodic oscillatory behavior.

\end{document}